\documentclass[preprint,authoryear,12pt]{elsarticle}

\usepackage{times}
\usepackage{bm}
\usepackage{natbib}
\usepackage{amsmath,amsfonts}
\usepackage{dsfont}
\usepackage{graphicx}

\usepackage{amssymb}

\journal{Computational Statistics $\&$ Data Analysis}

\begin{document}

\begin{frontmatter}



\title{Mixed Beta Regression: A Bayesian Perspective}


\author{Jorge I. Figueroa-Z\'u\~niga}
\address{Department of Statistics, Universidad de Concepci\'on, Santiago, Chile}
\author{Reinaldo B. Arellano-Valle}
\address{Department of Statistics, Pontificia Universidad Cat\'olica de Chile, Santiago, Chile}
\author{Silvia L.P. Ferrari}
\address{Department of Statistics, Universidade de S\~ao Paulo, Brazil}

\begin{abstract}
This paper builds on recent research that focuses on regression modeling of continuous bounded data,
such as proportions measured on a continuous scale. Specifically, it
deals with beta regression models with mixed effects from a Bayesian approach.
We use a suitable parameterization of the beta law in terms of its mean and a precision
parameter, and allow both parameters to be modeled through regression structures that
may involve fixed and random effects. Specification of prior distributions is discussed,
computational implementation via Gibbs sampling is provided, and illustrative examples are presented.
\end{abstract}


\end{frontmatter}


\section{Introduction}


Mixed-effects models have been widely employed in statistical
analysis, main-ly in the area of health, and their study
has been primarily restricted to response variables with normal
or, at least, symmetrical distributions.
Such models are not appropriately applicable when the response
variable has a support restricted to a doubly bounded interval.
Limited-range variables are, however, common in practice;
for example, proportions are bounded between zero and one.
This paper proposes a Bayesian analysis for mixed-effects regression
models that are tailored for situations where the response variable is
measured on a continuous scale and is restricted to the unit interval
$(0,1)$. Situations where the response, say $y$, is limited to a known
interval $(a,b)$ is also accommodated through the transformation
$y^{*}=(y-a)/(b-a)$. The response variable is assumed to be beta distributed with
mean (and possibly a precision parameter) modeled using fixed and random
effects. The substantial advantage to consider a beta modeling is due to
the flexibility that it provides. In fact, the beta family includes left
or right skewed, symmetric, J-shaped, and inverted J-shaped distributions.

Following \cite{FC04}, our proposed model uses a parameterization
of the beta law in terms of its mean and an additional positive
parameter that can be regarded as a precision parameter.
The mean of the response variable is conveniently linked
with a mixed-effects regression structure by the logit link function.
An extended version of such a model is also considered. It
assumes that the precision parameter is not constant over the
observations, but rather it is related to a mixed-effects function
through a log link.

To formulate our proposed models, we adopt a Bayesian approach.
We address the issues of model fitting via Gibbs sampling, choice
of prior distributions, and model selection based on the
deviance information criterion, the expected Akaike information criterion and the expected
Bayesian information criterion. Simulated and real data analysis
are presented for illustration. An appendix presents various pieces
of BUGS code used for fitting the mixed beta regression.


\section{Bayesian mixed beta regression}


Due to the flexibility of the beta distribution in terms of the
variety of density shapes that can be accommodated, this
distribution is a natural choice for modeling continuous data that
are restricted to the interval $(0,1)$. The probability density
function of a variable $y$ following a beta distribution
parameterized in terms of its mean $\mu$ ($0<\mu<1$) and a precision
parameter $\phi$ ($\phi>0$) is given by
\begin{eqnarray}\label{beta-density}
f(y|\mu,\phi) = \frac{\Gamma(\phi)}{\Gamma(\mu\phi)
\Gamma((1-\mu)\phi)}y^{\mu\phi-1}(1-y)^{(1-\mu)\phi-1}, ~~~~~
0<y<1,
\end{eqnarray}
where $\Gamma(\cdot)$ denotes the gamma function. Note that $\phi$ can be interpreted as a precision parameter,
since $\mu={\rm E}(y)$ and ${\rm Var}(y)=\mu(1-\mu)/(1+\phi)$
and, hence, for each fixed value of the mean $\mu$,
$1+\phi$ is inversely proportional to the variance of $y$. If
$y$ has density function (\ref{beta-density}), we write
$y\sim{\rm beta}(\mu\phi,(1-\mu)\phi)$.

Now, let $y_{1},\ldots,y_{n}$ be $n$ independent random variables such that\\
$y_{i}\sim{\rm beta}(\mu\phi,(1-\mu)\phi)$. The definition of a
beta regression model requires a transformation of the mean $\mu_i$ of $y_i$,
$i=1,\ldots,n$, that maps the interval $(0,1)$ onto
the real line. A convenient and popular link function is the
logit link. It is then assumed that
$\ln\{\mu_{i}/(1-\mu_{i})\}=x_{i}^\top\beta,$
where $x_{i}$ is a vector of known covariates for the $i$-th subject
and $\beta$ denotes a vector of regression coefficients.
The first element of $x_{i}$ is usually taken as 1 to allow for
an intercept.

The precision parameter $\phi$ may be assumed to be
constant over observations \citep{FC04} or it may be
modeled in terms of a regression structure \citep{SV06}.
Since the precision parameter is strictly positive, the log link function is a
natural choice. It is then assumed that $\ln(\phi_{i})=w_{i}^\top\delta$
where $w_{i}$ is a vector of covariates and $\delta$ denotes a vector of unknown
regression coefficients. Again, it is convenient to take the first element of $w_{i}$
as 1 to allow for an intercept in the precision description. There is no restriction
on whether or not the $w_{i}$s contain the same predictor variables as $x_{i}$s.

The beta regression model described above does not involve random effects.
Extending previous works on Bayesian generalized linear models
\citep{DGM00} and Bayesian beta regression \citep{BJT07}, we define below
two mixed beta regression models, the first of which assumes that the precision
parameter is the same for all the observations, and the second involves a
mixed-effects model for the precision parameter.

Let $y_1,\ldots,y_m$ be independent continuous random vectors, where\\
$y_{i}=(y_{i1},\ldots,y_{in_{i}})^\top$ represents  an observed response vector
for a sample unit $i$ and for which each of its components, $y_{ij}$, takes values on
the interval $(0,1)$. Consider also a regression model with the following structure:
\begin{eqnarray}\label{G}
G({\rm E}(y_{i}|b_{i}))=X_{i}\beta+Z_{i}b_{i},
\end{eqnarray}
$i=1,\ldots,m$, where $G(\cdot)$ is a vector-function linking the conditional mean
response vector ${\rm E}(y_{i}|b_{i})$ with the linear mixed model $\eta_{i}=X_{i}\beta+Z_{i}b_{i}$,
for which $X_i$ is the design matrix of dimension $n_i\times p$ corresponding to the vector
$\beta=(\beta_1,\ldots,\beta_p)^\top$ of regression coefficients (the fixed effects) and
$Z_{i}$ is the design matrix of dimension $n_i\times q$ associated with the vector
$b_{i}=(b_{i1},\ldots,b_{iq})^\top$ (the random effects).

For the logit link function, the $j$-th component of (\ref{G}) is
\begin{eqnarray*} \label{logit:mlm}
\ln\left\{\frac{\mu_{ij}}{1-\mu_{ij}}\right\}
=\eta_{ij}
=x_{ij}^\top\beta+z_{ij}^\top b_{i},
\end{eqnarray*}
where $\mu_{ij}={\rm E}(y_{ij}|b_i)$, $x_{ij}=(x_{ij1},\ldots,x_{ijp})^\top$, and
$z_{ij}=(z_{ij1},\ldots,z_{ijq})^\top$, which is equivalent to
\begin{eqnarray}\label{mu:mlm}
\mu_{ij}
=\frac{\exp(\eta_{ij})}{1+\exp(\eta_{ij})}
=\frac{\exp(x_{ij}^\top\beta+z_{ij}^\top b_{i})}{1+\exp(x_{ij}^\top\beta+z_{ij}^\top b_{i})}.
\end{eqnarray}

In this work, we first assume that for $i=1,2,\ldots,m$ and $j=1,2,\ldots,n_{i}$
$$y_{ij}|b_{i},\beta, \phi \stackrel{ind.}{\sim} {\rm beta}(\mu_{ij}\phi,(1-\mu_{ij})\phi),$$
i.e., conditionally on $b_{i}$, $\beta$, and $\phi$, the $y_{ij}$'s are independent and have
probability density function given by (\ref{beta-density}), with $\mu$ replaced by $\mu_{ij}$,
which is specified by (\ref{mu:mlm}). Note that in this formulation, $\phi$ represents a
common precision parameter.

In mixed models, the random effects $b_{1},\ldots,b_{m}$  are typically assumed to
be independent and normally distributed, namely $b_i|{\Sigma_{b}}{\stackrel{ind.}{\sim}}
N_q(0,\Sigma_{b})$, $i=1,\ldots,m$, where $\Sigma_b$ is a positive-definite
matrix. The normality assumption, however, can be inappropriate in practical applications
where the measurements present outliers. In these cases, it is more adequate to consider
multivariate distributions with heavier-than-normal tails for the random effects.
Consequently, a multivariate $t$-distribution with $\nu_b>0$ degrees of freedom, location
vector $\mu_b=0 \in\mathbb{R}^{q}$ and  positive-definite dispersion matrix $\Sigma_b$ is
a better candidate to model the random effects $b_{i}$'s, i.e., $b_{i}|\nu_{b},
{\Sigma_{b}}{\stackrel{ind.}{\sim}} t_{q}(\nu_{b},0,\Sigma_{b})$, $i=1,\ldots,m$. It should
be noticed here that for large values of $\nu_{b}$ the multivariate $t$-distribution is
approximately a multivariate normal distribution.

In the mixed beta regression model proposed above, the precision parameter $\phi$ is constant over the
observations. For a more general formulation of this model, we consider a different
precision parameter, say $\phi_{ij}$, for each response $y_{ij}$. We then assume a mixed
linear model for the logarithm of $\phi_{ij}$, namely
\begin{equation}
\ln(\phi_{ij})=\tau_{ij}=w_{ij}^\top\delta+h_{ij}^\top d_{i},
\label{phi:mlm}
\end{equation}
where $w_{ij}^\top=(w_{ij1},\ldots,w_{ijp^{*}})$  is the design vector corresponding to the
$p^*\times 1$ vector $\delta$ of fixed effects  and $h_{ij}^\top=(h_{ij1},\ldots,h_{ijq^{*}})$
is the design vector corresponding to the $q^{*}\times 1$ vector $d_{i}$ of random effects.
Note that the design matrices $W_{i}=(w_{i1},\ldots,w_{in_i})^\top$ and
$H_{i}=(h_{i1},\ldots,h_{in_i})^\top$ may, but are not required to, contain the same predictor
variables as the matrices $X_{i}=(x_{i1},\ldots,x_{in_i})^\top$ and $Z_{i}=(z_{i1},\ldots,z_{in_i})^\top$,
respectively. Here, it may be assumed that $d_i|{\Sigma_{d}}{\stackrel{ind.}{\sim}} N_q(0,\Sigma_{d})$,
$i=1,\ldots,m$, where $\Sigma_d$ is a positive-definite matrix. Alternatively, we may assume that
$d_{i}|\nu_{d},{\Sigma_{d}}{\stackrel{ind.}{\sim}} t_{q}(\nu_{d},0,\Sigma_{d})$, $i=1,\ldots,m$.

In order to complete the Bayesian specification of the beta mixed models described above,
elicitation of prior distributions for all unknown parameters is required. Multivariate normal
prior distributions are typically considered for the fixed effects, i.e.,
$\beta\sim  N_{p}(\mu_{\beta},\Sigma_{\beta})$.
Vague priors are usually specified by taking large values for the prior variances. However,
the impact of the scale choice under the normal model cannot be neglected. An alternative
strategy is to consider a multivariate $t$-distribution,
i.e., $\beta\sim t_{p}(\nu_{\beta},\mu_{\beta},\Sigma_{\beta})$ and to specify an appropriated value for
$\nu_{\beta}$, the degrees of freedom parameter.
If the vector of random effects is assumed to follow a multivariate $t$-distributed, i.e.,
$b_{i}|\nu_{b},\mu_{b},\Sigma_{b} \sim t_{q}(\nu_{b},0,\Sigma_{b})$, then the prior distribution
for the degrees of freedom can be discrete as in \cite{ACH93} and \cite{BGH95}, or continuous as
in \cite{G93}. We have chosen the latter alternative. More specifically, we consider an exponential
prior distribution with mean $1/a$ for the degrees of freedom, say $\varepsilon(a)$.
The prior distribution for the scale matrix of
random effects $\Sigma_{b}$ is chosen, mainly for computational simplicity, to be
an inverted Wishart distribution as in \cite{FRW10}, i.e., $\Sigma_{b}\sim IW_q(\psi,c)$.
An alternative prior distribution for $\Sigma_{b}$ is a constrained Wishart distribution
\citep{PC00}.

We now turn to the specification of prior distributions for the precision parameter.
As mentioned above, in this paper we study the following beta mixed regression models.\\

\noindent\emph{Model 1}: It considers the mixed regression model (\ref{mu:mlm}) for the
location parameters $\mu_{ij}$ and a common precision parameter $\phi$ for each observation $y_{ij}$.
In the Bayesian context, a natural choice for the prior distribution of the precision parameter would
be an inverse gamma distribution. If a slightly informative prior is required, it can be assumed that
$\phi \sim IG(\epsilon,\epsilon)$, with a small fixed positive value for $\epsilon$. \cite{G06} suggests
that the prior distribution $\phi=\texttt{U}^2$ with $\texttt{U} \sim U(0,a)$ with large $a$ ($a=50$
for example) is less informative than an inverse gamma prior. Here, we propose a more flexible prior
distribution for $\phi$ that includes Gelman's prior distribution as a special case. More specifically,
we propose the following prior specification for $\phi$: $\phi=(a B)^{2}$, where $B\sim
{\rm beta}(1+\epsilon,1+\epsilon)$, for  given positive values for $a$ and $\epsilon$.\\

\noindent\emph{Model 2}: It considers the mixed regression model (\ref{mu:mlm}) for the location
parameter $\mu_{ij}$  and a different precision parameter $\phi_{ij}$ for each $y_{ij}$, where
$\phi_{ij}$ is modeled as in (\ref{phi:mlm}). Here, the specification of prior distributions
for $\delta$ and the parameters of the distribution of the $d_{i}$s is similar to that
used for $\beta$ and the parameters of the distribution of the $b_{i}$s.


\section{Model fitting using Markov chain Monte Carlo sampling}


Let $y^\top=(y_1^\top,\ldots,y_m^\top)$ and $\eta^\top=(\eta_1^\top,\ldots,\eta_m^\top)$, where $\eta_i^\top=(\eta_{i1},\ldots,\eta_{in_i})$. Note that,
by assumption, conditionally on $\beta$, $\Sigma_{b}$, and $\nu_{b}$, the $\eta_i$s are independent and have density function
$f(\eta_{i}|\beta,\Sigma_{b},\nu_{b}) \propto f(b_{i}|\beta,\Sigma_{b},\nu_{b})$,
$i=1,\ldots,m$. We now present the following results for the joint posterior distribution under models 1 and 2 described in the previous section.

Under model 1, and the assumption that the parameters $\Sigma_{b}$, $\nu_{b}$, $\phi$, and $\beta$ are independent, the joint posterior density is
\begin{eqnarray*}
f(\beta,\Sigma_{b},\nu_{b},\phi,\eta|y) &\propto&
\left[\prod_{i=1}^{m}\prod_{j=1}^{n_{i}}f(y_{ij}|\eta_{ij},\phi)\right]\\
&\times&\left[\prod_{i=1}^{m}f(\eta_{i}|\beta,\Sigma_{b},\nu_{b})\right]
f(\Sigma_{b})f(\nu_{b})f(\phi)f(\beta).
\end{eqnarray*}
Gibbs sampling can be used to generate a Monte Carlo sample from the joint posterior density,
$f(\beta,\Sigma_{b},\nu_{b},\phi,\eta|y)$. The Gibbs sampler in this context
involves iteratively sampling from the full conditional
distributions:
$$f(\Sigma_{b}|\nu_{b},\beta,\phi,\eta,y), \ \ \ f(\nu_{b}|\Sigma_{b},\beta,\phi,\eta,y), \ \ \
f(\beta|\Sigma_{b},\nu_{b},\phi,\eta,y),$$
$$f(\phi|\beta,\Sigma_{b},\nu_{b},\eta,y), \ \ {\rm and} \ \
f(\eta_{i}|\eta_{k},\beta,\Sigma_{b},\nu_{b},\phi,y_i), \ \ i,\,k=1,\ldots,m, \ \ i \neq k,$$
which can be implemented in the WinBUGS software. Posterior inferences
on $\beta$, $\Sigma_{b}$, and $\phi$ and, more importantly, on the mean
responses $(\mu_{ij}; i=1,\dots,m, j=1,\ldots,n_i)$ are readily
obtained in WinBUGS. Hypothesis testing regarding regression
coefficients and mean responses are also straightforward.
\\

We now turn to model 2. Let $\tau^\top=(\tau_1^\top,\ldots,\tau_m^\top)$, where
$\tau_i^\top=(\tau_{i1},\ldots,\tau_{in_i})$. By assumption, conditionally on $\delta$, $\Sigma_{d}$,
and $\nu_{d}$, the $\tau_i$s are independent and have density function
$f(\tau_{i}|\delta,\Sigma_{d},\nu_{d}) \propto f(d_{i}|\delta,\Sigma_{d},\nu_{d})$,
$i=1,\ldots,m$. Assuming prior independence of $\Sigma_{b}$, $\nu_{b}$, $\Sigma_{d}$, $\nu_{d}$, $\delta$, and $\beta$,
we obtain the posterior density given by
\begin{eqnarray*}
f(\beta,\Sigma_{b},\nu_{b},\delta,\Sigma_{d},\nu_{d},\eta,\tau|y)
&\propto&
\left[\prod_{i=1}^{m}\prod_{j=1}^{n_{i}}f(y_{ij}|\eta_{ij},\tau_{ij})\right]\\
&\times&
\left[\prod_{i=1}^{m}f(\eta_{i}|\beta,\Sigma_{b},\nu_{b})\right]
\left[\prod_{i=1}^{m}f(\tau_{i}|\delta,\Sigma_{d},\nu_{d})\right]\\
&\times&
f(\Sigma_{b})f(\nu_{b})f(\delta)f(\Sigma_{d})f(\nu_{d})f(\beta).
\end{eqnarray*}
Similarly to model 1, the Gibbs sampling can be used to generate a Monte Carlo sample from
$f(\beta,\Sigma_{b},\nu_{b},\delta,\Sigma_{d},\nu_{d},\eta,\tau|y)$. In this case, the Gibbs sampler involves iteratively
sampling from the following full conditional distributions:
$$f(\Sigma_{b}|\nu_{b},\beta,\delta,\Sigma_{d},\nu_{d},\eta,\tau,y), \ \ \
f(\nu_{b}|\Sigma_{b},\beta,\delta,\Sigma_{d},\nu_{d},\eta,\tau,y),$$
$$f(\beta|\Sigma_{b},\nu_{b},\delta,\Sigma_{d},\nu_{d},\eta,\tau,y), \ \ \
f(\delta|\beta,\Sigma_{b},\nu_{b},\Sigma_{d},\nu_{d},\eta,\tau,y),$$
$$f(\Sigma_{d}|\beta,\Sigma_{b},\nu_{b},\delta,\nu_{d},\eta,\tau,y), \ \ \
f(\nu_{d}|\beta,\Sigma_{b},\nu_{b},\delta,\Sigma_{d},\eta,\tau,y),$$
$$f(\eta_{i}|\eta_{k},\tau_{k},\beta,\Sigma_{b},\nu_{b},\delta,\Sigma_{d},\nu_{d},y_i), \ \ {\rm and} \ \
f(\tau_{i}|\tau_{k},\eta_{k},\beta,\Sigma_{b},\nu_{b},\delta,\Sigma_{d},\nu_{d},y_i),$$
for $i,\,k=1,\ldots,m,$ and $i \neq k,$
which can be also implemented in the WinBUGS software. Thus, posterior inferences
on $\beta$, $\Sigma_{b}$, and $\phi_{ij}$, for $i=1,\dots,m,$ $j=1,\ldots,n_i$, and on
the mean responses $\mu_{ij}$, for  $i=1,\dots,m,$ $j=1,\ldots,n_i$ are easily
obtained in WinBUGS. Again, hypothesis testing regarding regression
coefficients and mean responses are also straightforward.


\section{Illustration via simulations}


$^{^{^{.}}}$\\
To illustrate the proposed methodology, we consider the following mixed beta regression model with
simulated data (model 1):
\begin{eqnarray*}
y_{ij}|b_{i},\phi,\beta&\sim& {\rm beta}(\mu_{ij}\phi,(1-\mu_{ij})\phi),
\end{eqnarray*}
where $\beta=(\beta_{1},\beta_{2},\beta_{3})^\top$, $b_{i}=(b_{i1},b_{i2})^\top$,
\begin{eqnarray*}
\ln\left\{\frac{\mu_{ij}}{1-\mu_{ij}}\right\}=\eta_{ij}= (\beta_{1} + b_{i1}) +
(\beta_{2}+b_{i2}) x_{ij2} + \beta_{3}x_{ij3},
\end{eqnarray*}
$i=1,\ldots,100$, $j=1,\ldots,5$,  and
$b_{i}|\nu_{b},\!\Sigma_{b}\sim t_{2}(\nu_{b},0,\Sigma_{b}).$
For our simulation study, the values of the covariates were generated from a uniform
distribution in the unit interval, and we set $\nu_{b}=10$, $\phi=49$, $\beta=(-2,1,2)^\top$,
and
\begin{eqnarray*}
\Sigma_{b}&=&\left(
\begin{array}{rr}
1 & ~ -0.3 \\
-0.3 & ~0.2 \\
\end{array}
\right).\\
\end{eqnarray*}

As proposed in Section 2, we adopt the following prior specifications: $\nu_{b}\sim\varepsilon(a)$,
$\Sigma_{b}\sim IW_2(\Psi,c)$, and $\beta=(\beta_1,\beta_2,\beta_3)^\top\sim t_3(\nu_\beta,\mu_\beta,\Sigma_\beta)$ with
$a=0.1,$ $c=5,$
{
$$\Psi=\left(
\begin{array}{rr}
20 & ~ 0 \\
0 & ~ 20  \\
\end{array}
\right),
\; \nu_\beta=10,
\; \mu_\beta=(0,0,0)^\top,
\; \Sigma_\beta=\left(
\begin{array}{rrr}
10 & ~ 0 & ~ 0 \\
0 & ~ 10 & ~ 0 \\
0 & ~ 0 & ~ 10 \\
\end{array}
\right).$$
}

We first analyze a single simulated dataset under different prior specifications for the precision parameter.
The following prior distributions
for $\phi$ were considered:
(i) $\phi \sim IG(\epsilon,\epsilon)$, with $\epsilon=0.001$ (model 1a);
(ii) $\phi=\texttt{U}^2$, with $\texttt{U} \sim U(0,50)$ (model 1b);
(iii) $\phi=(50 B)^{2}$ where $B \sim {\rm beta}(1+\epsilon,1+\epsilon)$, with $\epsilon=0.1$
(model 1c) and $0.5$ (model 1d);
(iv) $\ln(\phi)\sim t(\nu_{\beta},\mu_{\beta},\sigma_{\beta}^{2})$ with $\nu_{\beta}=10, \mu_{\beta}=0,
\sigma_{\beta}^{2}=10$ (model 1e). Note that (iii) corresponds to our proposal
(see Section 2). A sensitivity analysis for prior specification of the precision parameter can
be carried out from the figures in Table 1, which reports the deviance information criterion (DIC) proposed
by \cite{SBC02}, the expected Akaike information criterion (EAIC) introduced by \cite{B02},
and the expected Bayesian information criterion (EBIC) given in \cite{CL01}
for the fitted models with different prior distributions for $\phi$. We observe that
the different proposed priors lead to similar DICs, EAICs and EAICs. However, the three criteria indicate that
model 1d shows a slightly better fit than the other proposals.

\begin{table}[h!] \label{table1-DIC-model1}
\caption{\em DIC, EAIC and EBIC for the fitted models with different prior specifications for the precision parameter under model 1; simulated dataset}
\vspace{.3cm}
\small
 \begin{tabular*}{\textwidth}{@{\extracolsep{\fill}}ccccc}
 \hline
 Model        & Prior for $\phi$    & DIC  & EAIC & EBIC  \\
\hline
  model 1a & $\phi \sim IG(0.01,0.01)$ & $-1279.02$ & $-1411.84$ & $-1378.12$   \\[5pt]
  model 1b & $\phi={\rm U}^2, {\rm U}\sim U(0,50)$      & $-1279.09$  & $-1411.92$ & $-1378.2$  \\[5pt]
  model 1c & $\phi=(50 B)^{2}, B\sim {\rm beta}(1.1,1.1)$ & $-1279.26$  & $-1412.10$ & $-1378.38$  \\[5pt]
  model 1d & $\phi=(50 B)^{2}, B\sim {\rm beta}(1.5,1.5)$ & $-1279.84$   & $-1412.69$ & $-1378.98$  \\[5pt]
  model 1e & $\ln(\phi)\sim t(10,0,5)$ & $-1275.52$       & $-1406.92$ & $-1373.20$                          \\[5pt]
\hline
\end{tabular*}
\end{table}

In Table 2, we report the parameter estimates for model 1d.
These results show that the estimated parameters from
the Bayesian methodology proposed here are similar to the true
values of the model parameters. In our simulation study, we
consider 100,000 Monte Carlo iterations and the results are
presented considering the last 90,000 iterations.
In addition, the necessary diagnostic tests  (such as convergence, autocorrelation, history)
were performed, from which desirable behaviors were observed in
the chains (for brevity detailed numerical results are not shown but are commented below).
We also conducted a sensitivity analysis with respect to the prior specifications of the regression coefficients and the
dispersion matrix of the random effects coefficients. In each
case, the posterior inferences were not appreciably altered in
comparison with the results presented in Table 2.

\begin{table}[h!] \label{table2-estimates-model1d}
\caption{\em True mean and estimated posterior medians and means, $95\%$ credibility intervals (CI) for model 1d;
simulated dataset}
\small
\vspace{.3cm}
 \begin{tabular*}{\textwidth}{@{\extracolsep{\fill}}rrrrrr} \hline
\multicolumn{1}{c}{Parameter} & \multicolumn{5}{c}{Posterior Inference}\\
\cline{2-6}
                  & True    & Mean    & MC Error  & Median  & 95\% CI \\ \hline \\[-1pt]
 $\beta_{1}$      &$-2.000$       & $-2.094$    & $0.003$   & $-2.094$   &$(-2.329,\; -1.865)$\\[5pt]
 $\beta_{2}$      &$1.000$        & $1.074$     & $0.001$   & $1.075$    &$(0.906,\; 1.241)$\\[5pt]
 $\beta_{3}$      &$2.000$        & $2.000$     & $0.000$   & $2.000$    &$(1.873,\ 2.126)$\\[5pt]
 $\phi$           &$49.000$       & $49.280$    & $0.038$   & $49.150$   &$(41.350,\; 57.800)$\\[5pt]
 $\nu_{b}$	      &10.000           &7.086	    &0.058	    &5.338      &(2.223,23.100)\\[5pt]
 $\Sigma_{b_{11}}$&$1.000$        & $0.883$     & $0.002$   & $0.867$    &$(0.490,\; 1.369)$\\[5pt]
 $\Sigma_{b_{12}}$&$-0.300$       & $-0.182$    & $0.000$   & $-0.173$   &$(-0.393,\ -0.024)$\\[5pt]
 $\Sigma_{b_{22}}$&$0.200$        & $0.242$     & $0.001$   & $0.231$    &$(0.082,\; 0.466)$\\[5pt]
\hline
\end{tabular*}
\end{table}

The multivariate version of Gelman and Rubin's convergence diagnostic proposed by
\cite{BG98} indicates that the chain is convergent since the multivariate
proportional scale reduction factor (mprf) equals 1.01. Also, for each parameter, we checked
that the convergence is achieved for each chain. The latter
conclusion is corroborated by three different convergence tests, namely Gelman and Rubin's convergence
diagnostic \citep{GR92}, Geweke's diagnostic \citep{G92}, and Heidelberg and Welch's diagnostic
(\cite{HW81} and \cite{HW83}), which were obtained using the libraries {\tt lattice}
and {\tt coda} \citep{PB06} of the  {\tt R} sofware (freely available from
http://www.r-project.org/). To obtain  Gelman and Rubin's convergence diagnostic, we started two
chains in different initial points and performed $100,000$ Monte Carlo iterations, considering the last $90,000$ iterations. In addition, history and autocorrelation plots (not shown) suggest that the chain for each parameter is stationary  and not correlated, respectively. These results are essential to achieve an adequate estimation of the parameters.

We now turn to a simulation study in which we investigate the convenience of assuming a multivariate $t$ distribution
for the random effects. We consider different values for $\nu_{b}$ ($\nu_{b}=5, \ 10$ and $50$).
For each value of $\nu_{b}$, we generate $N=100$ datasets from the mixed beta regression model 1d (see above; the same values for the parameters and
sample size are used).
For each sample we fit the model under the assumption of multivariate $t$ and multiavariate normal distributed
random effects. We compute the bias  and the  root-mean-square error ($\sqrt{{\rm MSE}}$) for each
parameter estimator over the $N$ samples under the different settings.
Table 3 presents summary results for the estimation of all the parameters.
Also, for each sample, we compute the information criteria DIC, EAIC and EBIC for both fits.
Table 4 presents the mean DIC, EAIC and EBIC over the simulated samples.

Overall, figures in Table 3 suggest that, when the data are heavy-tailed distributed (say $\nu_b=5, 10$), the
performance of the posterior estimates obtained from the fit of the model that assumes a multivariate $t$ distribution for the
random effects is better than that of the posterior estimates taken from the normal fit.
From Table 4, advantage of the multivariate $t$ specification for the random effects over the
normal specification is clear, more so when  $\nu_b$ is small.

\begin{table}[h!] \label{table:table3-simulation}
\caption{\em Summary results based on 100 simulated datasets; t and normal fits}
\vspace{.3cm}
\scriptsize
\begin{tabular*}{\textwidth}{@{\extracolsep{\fill}}rrrrrrrrrrr}
\cline{1-11}
\multicolumn{1}{c}{$\nu_{b}$} & Fit &  & \multicolumn{8}{c}{Posterior Inference}\\
\cline{4-11}
          &                       &           & $\beta_{1}$   & $\beta_{2}$   & $\beta_{3}$ & $\Sigma_{b_{11}}$  & $\Sigma_{b_{12}}$ & $\Sigma_{b_{22}}$  & $\phi$  & $\nu_{b}$\\
\cline{1-11} \\[-1pt]
 $5$ & $t$   & Bias               & $0.006$ &	$-0.031$ & $0.003$  & $-0.108$ &	$0.140$	& $0.119$	& $0.294$	& $2.209$
\\[5pt]
 $ $ &       & $\sqrt{{\rm MSE}}$ & $0.104$ &	$ 0.078$ & $0.062$  & $ 0.251$	& $0.177$ &	$0.161$	& $3.900$	& $4.528$

\\[5pt] \\
 $ $ & normal& Bias	              & $0.019$ &	$-0.036$ & $0.003$  & $0.386$  & $0.048$	& $0.292$	& $-0.460$ &
\\[5pt]
 $ $ &       & $\sqrt{{\rm MSE}}$ & $0.120$	& $ 0.088$ & $0.063$  & $0.477$  & $0.183$	& $0.329$	& $ 4.031$ &
\\[5pt] \cline{1-11}
 $10$& $t$   & Bias               & $0.015$ & $-0,016$ & $-0.005$	& $-0.134$ & $0.128$  &	$0.092$	& $-1,281$ & $0.074$
\\[5pt]
 $ $ &       & $\sqrt{{\rm MSE}}$ & $0.112$	& $0.092$ &	 $ 0.064$	& $ 0.229$ & $0.161$  &	$0.144$	& $ 4.285$ & $3.768$
\\[5pt]\\
 $ $ & normal& Bias	              & $0.016$	& $-0.016$ & $-0.005$ &	$ 0.172$ & $0.071$  & $0.192$ &	$-1,609$ &
\\[5pt]
 $ $ &       & $\sqrt{{\rm MSE}}$ & $0.114$	& $ 0.092$ & $ 0.065$	& $ 0.276$ & $0.146$ &	$0.241$	& $ 4.480$ &
\\[5pt]\cline{1-11}
 $50$&  $t$  & Bias               & $0.020$	& $-0.018$ & $-0.005$	& $-0.163$ & $0.145$ &	$0.065$ & $-2.062$ & $-17.930$
\\[5pt]
 $ $ &       & $\sqrt{{\rm MSE}}$ & $0.114$	& $ 0.085$ & $ 0.061$ & $ 0.235$ & $0.173$ &  $0.135$	& $ 4.814$ &$20.379$
\\[5pt]\\
 $ $ & normal& Bias 	            & $0.018$ & $-0.018$ & $-0.007$ &	$-0.045$ & $0.124$ &  $0.103$ &	$-2.121$ &
 \\[5pt]
 $ $ &       & $\sqrt{{\rm MSE}}$ & $0.116$ & $ 0.085$ & $ 0.061$	& $ 0.181$ & $0.162$ &	$0.172$ & $ 4.917$ &
 \\[5pt]\cline{1-11}
\end{tabular*}
\end{table}
\begin{table}[h!] \label{table:table4-DIC-simulation}
\caption{\em Mean DIC, EAIC and EBIC  based on 100 simulated datasets; t and normal fits}
\vspace{.3cm}
\begin{tabular*}{\textwidth}{@{\extracolsep{\fill}}rrrrr}
 \hline
{$\nu_{b}$} & Fit           & DIC             & EAIC              & EBIC \\
\hline
 $5$        & $t$           & $-1319.86$      &$-1456.27$         &$-1422.55$ \\[5pt]
 $ $        & normal        & $-1313.76$      &$-1450.40$         &$-1416.68$ \\[5pt]
 \hline
 $10$       & $t$           & $-1276.30$      &$-1408.81$         &$-1375.09$ \\[5pt]
 $ $        & normal        & $-1274.88$      &$-1406.58$         &$-1372.86$ \\[5pt]
 \hline
 $50$       & $t$           & $-1249.76$      &$-1375.84$         &$-1342.12$ \\[5pt]
 $ $        & normal        & $-1249.91$      &$-1375.59$         &$-1341.87$ \\[5pt]
\hline
\end{tabular*}
\end{table}

We now use the same set of simulated dataset as in the beginning of this section to fit model 2, with five different regression
structures for the precision parameter. Note that the true (unknown) model is a mixed beta regression model with
constant precision, and hence only model 2a corresponds to the true model.
Prior distributions for the parameters $\nu_b$, $\Sigma_b$ and $\beta$ are the same as those
proposed for model 1. Also, for model specifications that include random effects for the precision parameter,
we assume that the precision random effects $d_i$ have the same distribution as the
location random effects $b_i$, namely $t(\nu_b,0,\Sigma_b)$. Table 5 reports the  DIC, EAIC and EBIC for the five
fitted models using simulated data. We observe that the submodels for the precision parameter that do
not include random effects achieve the best fits to our data, models 2a, 2c and 2d being similarly good.
However, model 2a, the model under which the data were simulated and which is equivalent to model 1e,
provides a better fit than the other proposals. Therefore, the best fitted model agrees with the true model.

\begin{table}[h!] \label{table:table5-DIC-model2}
\caption{\em DIC, EAIC and EBIC for the fitted models with different specifications of the precision parameter (model 2);
simulated dataset}
\vspace{.3cm}
{\small
 \begin{tabular*}{\textwidth}{@{\extracolsep{\fill}}cccccc} \hline
 Model       & Precision Mixed Model  & DIC & EAIC & EBIC  \\
                &      $(\ln(\phi_{ij}))$     &        \\
 \hline
  model 2a  & $\delta_{1}$                                     &  $-1275.52$  & $-1406.92$ & $-1373.20$ \\[5pt]
  model 2b  & $\delta_{1}+d_{i1}$                              &  $-1268.37$  & $-1399.64$ & $-1365.92$ \\[5pt]
  model 2c  & $\delta_{1}+\delta_{3}x_{ij3}$ &  $-1274.96$     & $-1406.34$   & $-1372.62$   \\[5pt]
  model 2d  & $\delta_{1}+\delta_{2}x_{ij2}+\delta_{3}x_{ij3}$ & $-1273.85$   & $-1405.16$ & $-1371.44$  \\[5pt]
  model 2e  & $(\delta_{1}+d_{i1})+(\delta_{2}+d_{i2})x_{ij2}+\delta_{3}x_{ij3}$ &  $-1270.13$  & $-1401.42$ & $-1367.70$   \\[5pt]
\hline
\end{tabular*}}
{\scriptsize{\it Note:  Models 2a-2e assumes the same location sub-model, namely ${\rm logit}(\mu_{ij})$ = $(\beta_{1}+b_{i1})+(\beta_{2}+b_{i2})x_{ij2}+x_{ij3}\beta_{3}$.}
}
\end{table}

Table 6 reports the parameter estimates under model 2a. It can be seen that the estimates obtained through the Bayesian methodoly proposed here are similar to the corresponding true values of the parameters. To fit model 2a, we considered $100,000$
Monte Carlo iterations and the estimates were obtained using the last $90,000$ iterations.
We obtained ${\rm mprf}=1.00<1.2$, indicating that the chain is convergent. Diagnostic plots (not shown)
suggest that the chain for each parameter is not correlated and stationary, respectively. Hence,
our estimates are reliable.

\begin{table}[h!] \label{table:table6-estimates-moel2a}
\caption{\em True mean and estimated posterior medians and means, $95\%$ credibility intervals (CI) for model 2a;
simulated data}
\vspace{.3cm}
\small
 \begin{tabular*}{\textwidth}{@{\extracolsep{\fill}}rrrrrr} \hline
\multicolumn{1}{c}{Parameter} & \multicolumn{5}{c}{Posterior Estimation}\\ \cline{2-6}
                & True    & Mean  & MC Error    &Median            & 95\% CI \\ \hline
 $\beta_{1}$    &$-2.000$         & $-2.091$    &$0.003$           &$-2.091$     &$(-2.325,-1.852)$\\[5pt]
 $\beta_{2}$    &$1.000$          & $1.073$     &$0.001$           &$1.073$      &$(0.905,1.242)$\\[5pt]
 $\beta_{3}$    &$2.000$          & $1.999$     &$0.000$           &$1.999$      &$(1.872,2.128)$\\[5pt]
 $\delta_{1}$   &$3.892$          &$3.885$		  &$0.000$	         &$3.887$      &$(3.715,4.048)$\\[5pt]
 $\nu_{b}$	        &$10.000$     &$7.083$	    &$0.059$	         &$5.357$      &$(2.258,23.000)$\\[5pt]
 $\Sigma_{b_{11}}$   &$1.000$     &$0.882$      &$0.002$           &$0.866$      &$(0.497,1.374)$\\[5pt]
 $\Sigma_{b_{12}}$   &$-0.300$    &$-0.180$     &$0.000$           &$-0.171$     &$(-0.392,-0.023)$\\[5pt]
 $\Sigma_{b_{22}}$   &$0.200$     &$0.239$      &$0.001$           &$0.228$      &$(0.078,0.457)$\\[5pt]
\hline
\end{tabular*}
\end{table}


\section{A real data application}


We now consider the dataset reported by \cite{P56}.  The response variable is the proportion of crude oil
converted into gasoline after distillation and fractionation. The dataset contains 32 observations on the response
and on other variables. By sorting the data, it is clear that there are only 10 crudes involved. A potentially
useful covariate is the end point ($EP$), i.e., the temperature (in degrees Fahrenheit) at which all gasoline has
vaporized. \cite{FC04} fitted a beta regression model with constant precision to these data, in which the batches of crude oil
are treated as a fixed factor with ten levels and with a fixed slope for the end point. Instead, \cite{V00} suggested that
the batches should be viewed as a random factor. Graphical inspection of the data suggests that a location
submodel with random intercepts and a common slope may be suitable for the data.

At the outset, we consider a mixed beta regression model with a constant precision parameter (model 1). The
location submodel involves random intercepts and a common slope. Table 7 gives the  DIC, EAIC and EBIC for
the model fitting with different prior specifications for the precision parameter $\phi$. As before,
for the parameters $\nu_b$, $\Sigma_b$, and $\beta$ we considered the
prior distributions
$\nu_{b}\sim\varepsilon(a)$,
$\Sigma_{b}\sim IW_2(\Psi,c)$, and $\beta\sim t_2(\nu_\beta,\mu_\beta,\Sigma_\beta)$ with
{
$$a=0.1,
\; \Psi=\left(
\begin{array}{rr}
20 & ~ 0 \\
0 & ~ 20  \\
\end{array}
\right),
\; c=4,
\; \nu_\beta=10,
\; \mu_\beta=(0,0)^\top,
\; \Sigma_\beta=\left(
\begin{array}{rr}
10 & ~ 0  \\
0 & ~ 10  \\
\end{array}
\right).$$
}
It can be noticed that the different proposed priors provide similar DIC, EAIC and EBIC values. The smallest EAIC and EBIC values are
obtained by a beta prior with $\epsilon=0.1$ and the smallest DIC is reached by a beta prior with $\epsilon=0.5$.

\begin{table}[h!] \label{table:tablePrater1}
\caption{\em DIC, EAIC and EBIC for the fitted models with different prior specifications of the precision parameter under model 1;
Prater's data}
\vspace{.3cm}
\small{
 \begin{tabular*}{\textwidth}{@{\extracolsep{\fill}}cccccc}
\hline
 Model       & Prior for $\phi$    & DIC & EAIC & EBIC   \\
                     &             &           \\
\hline
  model 1.1  & $\phi \sim IG(0.01,0.01)$ & $-141.485$ & -138.853 & -128.593   \\[5pt]
  model 1.2  & $\phi=\texttt{U}^2, \texttt{U}\sim U(0,50)$ & $-141.228$  & $-139.598$ & $-129.338$  \\[5pt]
  model 1.3  & $\phi=(50 B)^{2}, B\sim {\rm beta}(1.1,1.1)$ & $-142.062$ & $-140.069$ & $-129.809$   \\[5pt]
  model 1.4  & $\phi=(50 B)^{2}, B\sim {\rm beta}(1.5,1.5)$ & $-142.025$ & $-140.120$ & $-129.860$   \\[5pt]
\hline
\end{tabular*}
}
\scriptsize{Note. Models 1.1-1.4 assumes the same location sub-model, namely ${\rm logit}(\mu_{ij})$ =
$(\beta_{1}+b_{i1})+\beta_{2}EP_{ij}.$}
\end{table}

We now assume that $\phi$ is not constant through the observations. Again, the location submodel
assumes random intercepts and a common slope. As in the simulation study,
we consider  that both random effects, the $b_i$s and $d_i$s, are identically distributed with
distribution $t(\nu,0,\Sigma)$ (so that $\nu_d=\nu_b=\nu$ and $\Sigma_b=\Sigma_d=\Sigma$ in our
previous notation). Table 8 gives the DIC, EIAC and EBIC values for the model fitting under
different precision submodels (models 2.1--2.6). Note that model 2.1 is the same as model 1, i.e., it
implies constant precision but with a different prior for $\phi$, namely $\ln(\phi)\sim t(10,0,5)$.
Tables 9 and 10 give the posterior estimates of the parameters associated with models 1.4 and 2.5, which provide the
best fits for constant and noncontant precision, respectively.
Between the constant precision model (model 1.4) and the variable precision model (model 2.5),
the DIC, EAIC and EBIC values suggest that the later is the best. It means that not only the location
submodel but also the precision submodel are affected by a random additive effect and the
end point ($EP$). Also, there is no evidence of association between the
random effects since zero belongs to the credibility interval for $\Sigma_{12}$. It can
also be noticed that the covariate $EP$ affects both the mean and the precision
of the proportion of crude oil converted into gasoline positively.

\begin{table}[h!] \label{table:tablePrater2}
\caption{\em DIC, EAIC and EBIC for the fitted models with different specifications for the precision parameter under model 2; Prater's data}
\small
\vspace{.3cm}
 \begin{tabular*}{\textwidth}{@{\extracolsep{\fill}}cccccc}
\hline
 Model        & Precision submodel & DIC &EAIC & EBIC\\
                        & $(\ln(\phi_{ij}))$       &           \\
     \hline
   model 2.1  & $\delta_{1}$                             & $-141.382$ &$-139.006$&$-128.746$  \\
   model 2.2  & $\delta_{1}+d_{i1}$                      & $-140.324$ &$-138.970$&$-128.716$   \\
   model 2.3  & $\delta_{2}EP_{ij}$                      & $-144.944$ &$-142.596$&$-132.336$  \\
   model 2.4  & $\delta_{1}+\delta_{2} EP_{ij}$          & $-144.219$ &$-142.239$&$-131.980$  \\
   model 2.5  & $d_{i1}+\delta_{2}EP_{ij}$               & $-146.026$ &$-145.086$&$-134.826$  \\
   model 2.6  & $(\delta_{1}+d_{i1})+\delta_{2}EP_{ij}$  & $-144.839$ &$-144.066$&$-133.806$ \\
\hline
\end{tabular*}
{\scriptsize{\it Note:  Models 2.1-2.6 assumes the same location sub-model, namely  \quad ${\rm logit}(\mu_{ij})$ =
$(\beta_{1}+b_{i1})+\beta_{2}EP_{ij}.$}}
\end{table}

\begin{table}[h!] \label{table:table9-estimates-model14}
\caption{\em Estimated posterior medians and means, $95\%$ credibility intervals (CI) for the mixed beta regression model 1.4;
Prater's data}
 {\footnotesize
\vspace{.3cm}
 \begin{tabular*}{\textwidth}{@{\extracolsep{\fill}}rrrrrr} \hline
\multicolumn{1}{c}{Parameter} & \multicolumn{4}{c}{Posterior Inference}\\
\cline{2-5}
                  & Mean    & MC Error  & Median  & 95\% CI \\ \hline
 $\beta_{1}$      & $-5.116$	                    &$0.004$     	&$-5.112$     &$(-5.631,-4.628)$\\[5pt]
 $\beta_{2}$      & $10.730\!\times\!10^{-3}$	    &$0.533\!\times\! 10^{-3}$	&$10.740\!\times\! 10^{-3}$   &($9.629\!\times\! 10^{-3}$,$11.760\!\times \!10^{-3})$\\[5pt]
 $\nu$            & $12.99$	                      &$0.222$	    &$9.334$      &$(1.098,45.72)$\\[5pt]
 $\phi$           & $296.1$		                    &$1.387$	    &$289.100$    &$(142.100,500.400)$\\[5pt]
 $\Sigma_{11}$    & $0.204$	                      &$0.002$	    &$0.175$      &$(0.041,0.519)$\\[5pt]
 $\Sigma_{12}$    & $0.464\!\times\! 10^{-3}$	    &$2.309\!\times\! 10^{-3}$	&$1.337\!\times\! 10^{-3}$     &$(-0.329,0.320)$\\[5pt]
 $\Sigma_{22}$    & $0.121$	                  &$0.005$	    &$0.044$    &$(0.007,0.703)$\\[5pt]
\hline
\end{tabular*}
}
\end{table}

\begin{table}[h!] \label{table:table10-estimates-model25}
\caption{\em Estimated posterior medians and means, $95\%$ credibility intervals (CI) for the mixed beta regression model 2.5;
Prater's data}
 {\scriptsize
\vspace{.3cm}
 \begin{tabular*}{\textwidth}{@{\extracolsep{\fill}}rrrrrr}
\cline{1-5}
\multicolumn{1}{c}{Parameter} & \multicolumn{4}{c}{Posterior Inference in location sub-model}\\
\cline{2-5}

                  & Mean    & MC Error  & Median  & 95\% CI \\
\cline{1-5}
 $\beta_{1}$      &$-4.783$	      &$0.008$	   &$-4.780$     &$(-5.348,\;-4.233)$\\[5pt]
 $\beta_{2}$      &$9.892\!\times\!10^{-3}$	 &$0.015\!\times\!10^{-3}$ &$9.898\!\times\!10^{-3}$ &($8.673\!\times\!10^{-3}$,\;$11.080\!\times\!10^{-3}$)\\[5pt]	
 $\delta_{2}$     &$17.150\!\times\!10^{-3}$	  &$0.008\!\times\!10^{-3}$	   &$17.190\!\times\!10^{-3}$    &($14.770\!\times\!10^{-3}$,\;$19.330\!\times\!10^{-3}$)     \\[5pt]  			
 $\nu$            &$13.190$	      &$0.066$	   &$9.333$      &$(1.343,\;47.01)$\\[5pt]
 $\Sigma_{11}$    &$0.179$	      &$0.000$	   &$0.158$      &$(0.039,\;0.444)$\\[5pt]
 $\Sigma_{12}$    &$-0.930\!\times\!10^{-3}$	      &$1.478\!\times\!10^{-3}$   &$0.358\!\times\!10^{-3}$     &$(-0.290\!,\;0.283$)\\[5pt]
 $\Sigma_{22}$    &$0.108$	      &$0.002$	   &$0.041$     &$(0.007,\;0.627)$\\[5pt]
\hline
\end{tabular*}
}
\end{table}

\begin{figure}[h]
\renewcommand{\figurename}{Figure}
\begin{center}
\includegraphics[width=4in,height=4.726in]{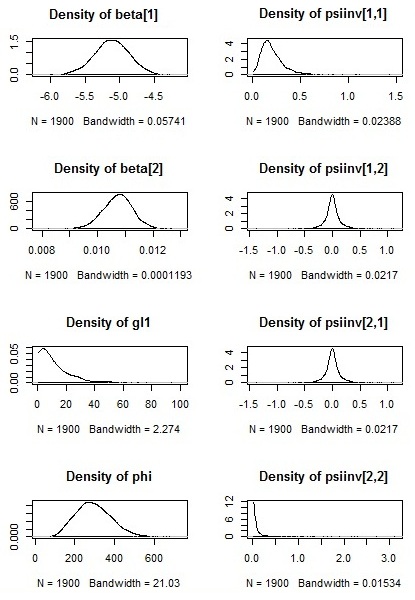}
\end{center}
\caption{\em Density; Prater's data; constant $\phi$; model 1.4
  \label{fig1}}
\end{figure}

\begin{figure}[h]
\renewcommand{\figurename}{Figure}
\begin{center}
\includegraphics[width=4in,height=4.726in]{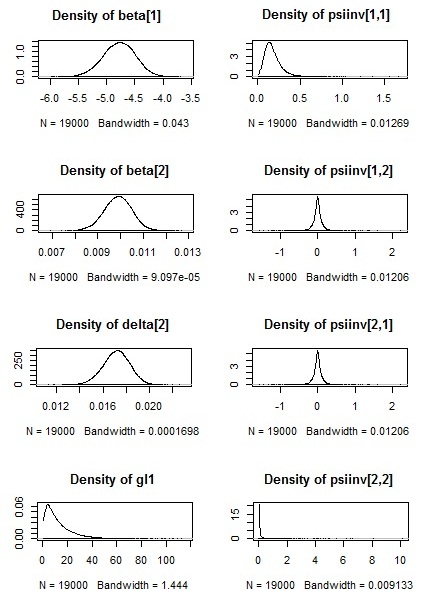}
\end{center}
\caption{\em Density; Prater's data; non constant $\phi$; model 2.5
  \label{fig2}}
\end{figure}

Some technical details relating to the fit of the models are now in order.
We considered $200,000$ Monte Carlo iterations and our results were obtained
considering the last $190,000$ iterations.
Additionally, we performed the diagnostic tests reported for the simulated data,
all of which suggested suitable behavior of the chains.
For model 1.4, the multivariate version of Gelman and Rubin's convergence diagnostic
\citep{BG98} indicates that the chain is convergent
(${\rm mprf}=1.09 < 1.2$). Also, diagnostic plots (not shown) suggest that the chain for each parameter
is not correlated and stationary, respectively, while Figure \ref{fig1}
demonstrates that the posterior densitity function for each parameter
does not present multimodality; it should be noted that multimodality
can be accompanied by convergence problems. We can then assume that the estimates reported
in Table 9 are reliable. For model 2.5, similar diagnostic evidence was obtained. Here,
${\rm mprf}=1.06$ and diagnostic plots (not shown) and Figure \ref{fig2} suggest that the results in Table 10
can be trusted.

\section{Discussion}
Beta regression modeling has gained increasing popularity after the work of \cite{FC04}, who described
a beta regression model parameterized in terms of the mean response and a common precision parameter,
 and developed frequentist inference and basic diagnostic tools for the proposed model.
A complementary approach proposed by \cite{SV06} considers  that the precision parameter is not fixed
but, instead, is modeled in a regression manner. A Bayesian beta regression model was studied by
\cite{BJT07}. In this paper, we extended these ideas for a mixed beta regression model under a
Bayesian perspective.

The present paper considered Bayesian inference for mixed beta regression based
on two different approaches. First, the precision parameter was assumed to be fixed, i.e., the same for all observations.
A linear regression structure was proposed for the mean parameter through a logit link function. Our results are
readily extended to other link choices.
Specification of different priors for the common precision parameter was studied. We considered a prior distribution
for $\phi$ of the type $\phi=\texttt{U}^{2}$, with $\texttt{U} \sim U(0,a)$, where it is common to consider as
initial value $a=50$  \citep{G06}. We also proposed alternative priors, namely, prior distributions of the type $\phi=(aB)^{2}$,
with $B \sim {\rm beta}(1+\epsilon,1+\epsilon)$ and  $\epsilon=0.001,0.01,0.1,0.5,\ldots$, which delivered good
results in terms of model fit and performance of diagnostics tests.
Second, the precision parameter was modeled through its own linear regression structure using a log link. Again, other choices
of the precision link function can be accommodated. For both the mean and the precision submodels, a mixed-effects model
with a multivariate $t$ distribution for the fixed and the random effects was considered. Our empirical applications yielded
good results in terms of model fit and diagnostic tests. It is worth mentioning that in this context, it is necessary to perform
a careful model selection for the precision modeling including more or fewer fixed and random effects since it is not clear in advance
which model is more plausible.

A classic version of this problem was raised by \cite{Z10}, where mixed beta regression models were estimated using the SAS procedure NLMIXED (\cite{SAS08}), employing adaptive Gaussian quadrature. This approach achieves good results, but the implementation of the mixed beta regression model for random effects that are non-normally distributed is very challenging. In this sense, our approach is more flexible because one can  easily implement it when the distribution of the  random effects follow a normal, Student-$t$, skew normal or another distribution, by using simple and accessible software such as WinBUGS. Another advantage of this approach is the easy implementation for the imputation of missing data \citep{CBDM07}, a common situation in practice and for which a classic approach is much more complicated.


\section*{Appendix: BUGS codes for the mixed beta regression}
This appendix presents the various pieces of BUGS code used for fitting the mixed beta regression in the simulated data example.\\

\medskip

{\scriptsize
\emph{Inverse gamma prior for $\phi$}
\begin{verbatim}
  model
    {
        for( i in 1 : m ) {
        for( j in 1 : n ) {
        Y[i , j] ~ dbeta(a1[i,j] ,a2[i,j])
        a1[i,j] <-  mu[i , j]*phi
        a2[i,j] <- (1-mu[i , j])*phi
        logit(mu[i , j]) <- inprod(x[i, j, ], beta[ ])+inprod(z[i, j, ], b[i,1,])
            }
        b[i,1,1:q ]  ~ dmt(cerovec [ ] ,psi[ , ],gl1)
                }
            gl1~dexp(a0)
        beta[1:p] ~ dmt(alpha[  ] , V1[ , ],gl2)
        V1[1:p ,1:p] <- inverse(V[ , ])
        psi[1:q,1:q] ~ dwish(R0[ , ], c0)
        psiinv[1:q,1:q]<-inverse(psi[1:q,1:q])
        phiinv ~ dgamma(a00,a00)
        phi<-1/phiinv
            }
\end{verbatim}

\emph{Uniform prior for $\phi$}
\begin{verbatim}
  model
    {
        for( i in 1 : m ) {
        for( j in 1 : n ) {
        Y[i , j] ~ dbeta(a1[i,j] ,a2[i,j])
        a1[i,j] <-  mu[i , j]*phi
        a2[i,j] <- (1-mu[i , j])*phi
        logit(mu[i , j]) <- inprod(x[i, j, ], beta[ ])+inprod(z[i, j, ], b[i,1,])
            }
        b[i,1,1:q ]  ~ dmt(cerovec [ ] ,psi[ , ],gl1)
                }
        gl1~dexp(a0)
        beta[1:p] ~ dmt(alpha[  ] , V1[ , ],gl2)
        V1[1:p ,1:p] <- inverse(V[ , ])
        psi[1:q,1:q] ~ dwish(R0[ , ], c0)
        psiinv[1:q,1:q]<-inverse(psi[1:q,1:q])

        phir ~ dunif(a00,b01)
        phi<- phir*phir
            }
\end{verbatim}

\emph{Beta prior for $\phi$}
\begin{verbatim}
  model
    {
        for( i in 1 : m ) {
        for( j in 1 : n ) {
        Y[i , j] ~ dbeta(a1[i,j] ,a2[i,j])
        a1[i,j] <-  mu[i , j]*phi
        a2[i,j] <- (1-mu[i , j])*phi
        logit(mu[i , j]) <- inprod(x[i, j, ], beta[ ])+inprod(z[i, j, ], b[i,1,])
            }
        b[i,1,1:q ]  ~ dmt(cerovec [ ] ,psi[ , ],gl1)
                }
        gl1~dexp(a0)
        beta[1:p] ~ dmt(alpha[  ] , V1[ , ],gl2)
        V1[1:p ,1:p] <- inverse(V[ , ])
        psi[1:q,1:q] ~ dwish(R0[ , ], c0)
        psiinv[1:q,1:q]<-inverse(psi[1:q,1:q])
        phiinicial ~ dbeta(a00,b0)
        phi<-(phiinicial*b11)*(phiinicial*b11)
            }
\end{verbatim}

\emph{Submodel for $\phi$}
\begin{verbatim}
  model
    {
        for( i in 1 : m ) {
        for( j in 1 : n ) {
        Y[i , j] ~ dbeta(a1[i,j] ,a2[i,j])
        a1[i,j] <-  mu[i , j]*phi[i,j]
        a2[i,j] <- (1-mu[i , j])*phi[i,j]
        log(phi[i,j])<-inprod(x[i, j, ], delta[ ])+inprod(z[i, j, ], gama[i,1,])
        logit(mu[i , j]) <- inprod(x[i, j, ], beta[ ])+inprod(z[i, j, ], b[i,1,])
            }
        b[i,1,1:q ]  ~ dmt(cerovec [ ] ,psi[ , ],gl1)
        gama[i,1,1:q ]  ~ dmt(cerovec [ ] ,psi[ , ],gl1)
                }
        gl1~dexp(a0)
        beta[1:p] ~ dmt(alpha[  ] , V1[ , ],gl2)
        delta[1:p] ~ dmt(alpha[  ] , V1[ , ],gl2)
        V1[1:p ,1:p] <- inverse(V[ , ])
        psi[1:q,1:q] ~ dwish(R0[ , ], c0)
        psiinv[1:q,1:q]<-inverse(psi[1:q,1:q])
        meanphi<-mean(phi[,])
            }
\end{verbatim}
}

\section*{Acknowledgements}

We thank both referees for their constructive comments and suggestions. This research was partially supported by grant FONDECYT 1120121-Chile, by Facultad de Matem\'aticas y Vicerrector\'ia de Investigaci\'on (VRI) of the Pontificia Universidad Cat\'olica de Chile, and by CNPq-Brazil.

\newpage 

\section*{References}
\bibliographystyle{elsarticle-harv}
\bibliography{XBib}

\end{document}